\documentclass[11pt]{article}
\usepackage{sigsam, amsmath}

\issue{}%Vol 40, No. 3, Issue 197, Month 2016}
\articlehead{}%ISSAC poster abstracts}

\titlehead{Satisfiability Checking and Symbolic Computation}
\authorhead{E.~\'{A}brah\'{a}m et al.
}

\usepackage{shortlst}
\usepackage{color}
\usepackage{tikz}
\usetikzlibrary{shadows}
\usetikzlibrary{positioning}
\usetikzlibrary{fit}
\usetikzlibrary{backgrounds}
\usetikzlibrary{chains}
\usetikzlibrary{matrix}
\usetikzlibrary{snakes}
\usetikzlibrary{shapes}
\usetikzlibrary{calc}
\usetikzlibrary{arrows}

\newcommand{\scsc}{\textsf{SC$^\mathsf{2}$}\xspace}
\usepackage{xspace}
\newcommand{\smtlib}{\textit{SMT-LIB}\xspace}

\newcommand{\smtrat}{{\small\texttt{SMT-RAT}}\xspace}

\newcommand{\altergo}{{\small\texttt{Alt-Ergo}}\xspace}
\newcommand{\opensmt}{{\small\texttt{OpenSMT2}}\xspace}
\newcommand{\cvc}{{\small\texttt{CVC4}}\xspace}
\newcommand{\yices}{{\small\texttt{Yices2}}\xspace}
\newcommand{\mathsat}{{\small\texttt{MathSAT}}\xspace}

\newcommand{\minismt}{{\small\texttt{MiniSmt}}\xspace}
\newcommand{\isat}{{\small\texttt{iSAT3}}\xspace}
\newcommand{\zthree}{{\small\texttt{Z3}}\xspace}

\newcommand{\verit}{{\small\texttt{veriT}}\xspace}

\begin{document}

\title{Satisfiability Checking and Symbolic Computation}

\author{E.~\'{A}brah\'{a}m$^{1}$,  
J.~Abbott$^{11}$,  
B.~Becker$^{2}$, 
A.M.~Bigatti$^{3}$,
M.~Brain$^{10}$, 
B.~Buchberger$^{4}$, \\
A.~Cimatti$^{5}$,  
J.H.~Davenport$^{6}$,  
M.~England$^{7}$,  
P.~Fontaine$^{8}$, \\
S.~Forrest$^{9}$,  
A.~Griggio$^{5}$,   
D.~Kroening$^{10}$,  
W.M.~Seiler$^{11}$ and
T.~Sturm$^{12}$ 
\\ \qquad \\
\small{
$^1$RWTH Aachen University, Aachen, Germany; \quad
$^2$Albert-Ludwigs-Universit\"{a}t, Freiburg, Germany; 
} \\
\small{
$^3$Universit\`{a} degli studi di Genova, Italy; \quad
$^4$Johannes Kepler Universit\"{a}t, Linz, Austria; 
} \\
\small{
$^5$Fondazione Bruno Kessler, Trento, Italy; \quad
$^6$University of Bath, Bath, U.K.; 
} \\
\small{
$^7$Coventry University, Coventry, U.K.; \quad
$^8$LORIA, Inria, Universit\'{e} de Lorraine, Nancy, France; 
} \\
\small{ $^9$Maplesoft Europe Ltd; \quad
$^{10}$University of Oxford, Oxford, U.K.; \quad
$^{11}$Universit\"{a}t Kassel, Kassel, Germany; 
} \\
\small{
$^{12}$CNRS, LORIA, Nancy, France and Max-Planck-Institut f\"{u}r Informatik, Saarbr\"{u}cken, Germany.
}
}

\date{} % leave this empty

\maketitle

\begin{abstract}
\emph{Symbolic Computation} and \emph{Satisfiability Checking} are viewed as individual research areas, but they share common interests in the development, implementation and application of decision procedures for arithmetic theories. Despite these commonalities, the two communities are currently only weakly connected.  We introduce a new project \scsc to build a joint community in this area, supported by a newly accepted EU (H2020-FETOPEN-CSA) project of the same name.  
We aim to strengthen the connection between these communities by creating common platforms, initiating interaction and exchange, identifying common challenges, and developing a common roadmap.  This abstract and accompanying poster describes the motivation and aims for the project, and reports on the first activities.
%and formalise some relevant challenges for the unified \scsc community.
\end{abstract}

\section{Introduction} 
\label{SEC:Intro}

We describe a new project to bring together the communities of \emph{\textbf{S}ymbolic \textbf{C}omputation} and \emph{\textbf{S}atisfiability \textbf{C}hecking} into a new joint community, \scsc.  
Both communities have long histories, as illustrated by the tool development timeline in Figure \ref{fig:tools}, but traditionally they do not interact much even though they are now individually addressing similar problems in non-linear algebra.  
In Section~\ref{SEC:SAT} we give an introduction to \emph{Satisfiability Checking} (a corresponding introduction to Symbolic Computation is omitted given the audience for this abstract).  
We then discuss some of the challenges for the new \scsc community in Section~\ref{SEC:Challenges} and the project actions planned to address them.  The reader is referred to \cite{SC2CICM} for more details and full references; and the \scsc website (\url{http://www.sc-square.org}) for new information as it occurs.
The accompanying poster is available at: \url{http://www.sc-square.org/SC2-AnnouncementPoster.pdf}.

%%%%% History Figure %%%%%

\definecolor{darkred}{rgb}{0.65,0,0}
\definecolor{darkblue}{rgb}{0,0,0.4}
\definecolor{owngreen}{rgb}{0,0.6,0}
\definecolor{darkgreen}{rgb}{0,0.35,0}
\definecolor{green}{rgb}{0,0.55,0}
\definecolor{gray}{rgb}{0.7,0.7,0.7}
\definecolor{darkgrey}{rgb}{0.7,0.7,0.7}
\definecolor{brown}{cmyk}{0, 0.8, 1, 0.6}

\newcommand{\cbb}[1]{{\color{blue}#1}}
\newcommand{\cdb}[1]{{\color{darkblue}#1}}
\newcommand{\cg}[1]{{\color{green}#1}}
\newcommand{\cdg}[1]{{\color{darkgreen}#1}}
\newcommand{\cred}[1]{{\color{red}#1}}
\newcommand{\cdred}[1]{{\color{darkred}#1}}
\newcommand{\cG}[1]{{\color{gray}#1}}

\begin{figure}[t]
\begin{center}
  \scalebox{0.85}{
    \begin{tikzpicture}
      \path[draw,thick,-latex'] (-0.1,0)--(12.3,0);
      \path[draw] (0,-0.1)--(0,0.1);
      \path[draw] (2,-0.1)--(2,0.1);
      \path[draw] (4,-0.1)--(4,0.1);
      \path[draw] (6,-0.1)--(6,0.1);
      \path[draw] (8,-0.1)--(8,0.1);
      \path[draw] (10,-0.1)--(10,0.1);
      \path[draw] (12,-0.1)--(12,0.1);
      \node at (0,-0.3) {\tiny 1960};
      \node at (2,-0.3) {\tiny 1970};
      \node at (4,-0.3) {\tiny 1980};
      \node at (6,-0.3) {\tiny 1990};
      \node at (8,-0.3) {\tiny 2000};
      \node at (10,-0.3) {\tiny 2010};
      \node at (12,-0.3) {\tiny 2020};
      
      \node at (0.5,2.5) {\large \cbb{CAS}};
      \node at (0.5,-2) {\large \cred{SAT}};
      \node at (0.5,-3) {\large \cg{SMT}};
      
      \node[rotate=60,anchor=west] at (0.6,0.25) {\scriptsize\tt\cbb{Schoonschip}};
      \node[rotate=60,anchor=west] at (1.0,0.25) {\scriptsize\tt\cbb{MATHLAB}};
      \node[rotate=60,anchor=west] at (1.6,0.25) {\scriptsize\tt\cbb{Reduce} \cdb{Altran}};
      \node[rotate=60,anchor=west] at (2.2,0.25) {\scriptsize\tt\cbb{Scratchpad/Axiom}};
      \node[rotate=60,anchor=west] at (3.4,0.25) {\scriptsize\tt\cbb{Macsyma}};
      \node[rotate=60,anchor=west] at (3.8,0.25) {\scriptsize\tt\cbb{SMP}};
      \node[rotate=60,anchor=west] at (4.2,0.25) {\scriptsize\tt\cbb{muMATH}};
      \node[rotate=60,anchor=west] at (4.6,0.25) {\scriptsize\tt\cbb{Maple}};
      \node[rotate=60,anchor=west] at (5.0,0.25) {\scriptsize\tt\cbb{Mathcad} \cdb{SAC} \cbb{GAP}};
      \node[rotate=60,anchor=west] at (5.4,0.25) {\scriptsize\tt\cbb{CoCoA} \cdb{MathHandbook} \cbb{Mathomatic}};
      \node[rotate=60,anchor=west] at (5.8,0.25) {\scriptsize\tt\cbb{Mathematica} \cdb{Derive} \cbb{FORM}};
      \node[rotate=60,anchor=west] at (6.2,0.25) {\scriptsize\tt\cbb{KASH/KANT} \cdb{PARI/GP} \cbb{Singular}};
      \node[rotate=60,anchor=west] at (6.8,0.25) {\scriptsize\tt\cbb{Magma} \cdb{Fermat} \cbb{Erable} \cdb{Macaulay2}};
      \node[rotate=60,anchor=west] at (7.2,0.25) {\scriptsize\tt \cdb{SymbolicC++}};
      \node[rotate=60,anchor=west] at (7.6,0.25) {\scriptsize\tt\cbb{Maxima}};
      \node[rotate=60,anchor=west] at (8,0.25) {\scriptsize\tt\cbb{Xcas/Giac}};
      \node[rotate=60,anchor=west] at (8.4,0.25) {\scriptsize\tt\cbb{Yacas}};
      \node[rotate=60,anchor=west] at (8.8,0.25) {\scriptsize\tt\cbb{SAGE} \cdb{SMath} \cbb{Studio}};
      \node[rotate=60,anchor=west] at (9.2,0.25) {\scriptsize\tt\cbb{Cadabra} \cdb{SymPy} \cbb{OpenAxiom}};
      \node[rotate=60,anchor=west] at (9.6,0.25) {\scriptsize\tt\cbb{MATLAB} \cdb{MuPAD}};
      \node[rotate=60,anchor=west] at (10.0,0.25) {\scriptsize\tt\cbb{Wolfram Alpha} \cdb{TI-Nspire} \cbb{CAS}};
      \node[rotate=60,anchor=west] at (10.5,0.25) {\scriptsize\tt\cbb{Mathics} \cdb{Symbolism} \cbb{FxSolver}};
      \node[rotate=60,anchor=west] at (11,0.25) {\scriptsize\tt\cbb{Calcinator} \cdb{SyMAT} \cbb{Mathemagix}};
    
    \node[rotate=-60,anchor=west] at (6.8,-0.4) {\scriptsize\tt\cred{WalkSAT} \cdred{SATO}};
    \node[rotate=-60,anchor=west] at (7.1,-0.4) {\scriptsize\tt\cg{Simplify} \cdg{SVC}};
    \node[rotate=-60,anchor=west] at (7.4,-0.4) {\scriptsize\tt\cred{GRASP} \cdred{Chaff} \cred{BCSAT}};
    \node[rotate=-60,anchor=west] at (7.8,-0.4) {\scriptsize\tt\cred{MiniSAT} \cdred{Berkmin} \cred{zChaff} \cdred{Siege}};
\node[rotate=-60,anchor=west] at (8.2,-0.4) {\scriptsize\tt \cg{ICS} \cdg{Uclid}  \cg{MathSAT} \cdg{Barcelogic} };
\node[rotate=-60,anchor=west] at (8.6,-0.4) {\scriptsize\tt\cred{HyperSat} \cdred{RSat} \cred{Sat4j}};
\node[rotate=-60,anchor=west] at (9.0,-0.4) {\scriptsize\tt\cg{Yices} \cdg{CVC} \cg{HySAT/iSAT} \cdg{DPT}};

\node[rotate=-60,anchor=west] at (9.3,-0.4) {\scriptsize\tt\cg{Z3} \cdg{Alt-Ergo} \cg{Beaver} \cdg{ABsolver} };
\node[rotate=-60,anchor=west] at (9.7,-0.4) {\scriptsize\tt\cg{Boolector} \cdg{PicoSAT} \cg{Spear}};
\node[rotate=-60,anchor=west] at (10.3,-0.4) {\scriptsize\tt\cg{MiniSmt} \cdg{veriT} \cg{OpenCog}};
\node[rotate=-60,anchor=west] at (10,-0.4) {\scriptsize\tt\cred{ArgoSat}};
\node[rotate=-60,anchor=west] at (10,-0.4) {\scriptsize\tt\phantom{ArgoSat} \cg{OpenSMT} \cdg{SatEEn} \cg{SWORD}};
\node[rotate=-60,anchor=west] at (10.6,-0.4) {\scriptsize\tt\cred{Glucose} \cdred{CryptoMiniSat}};
\node[rotate=-60,anchor=west] at (10.6,-0.4) {\scriptsize\tt\phantom{Glucose CryptoMiniSat} \cdg{SONOLAR}};
\node[rotate=-60,anchor=west] at (10.9,-0.4) {\scriptsize\tt\cred{Lingeling} \cdred{UBCSAT}};
\node[rotate=-60,anchor=west] at (10.9,-0.4) {\scriptsize\tt\phantom{Lingeling UBCSAT} \cg{SMTInterpol}};
\node[rotate=-60,anchor=west] at (11.2,-0.4) {\scriptsize\tt\cg{SMT-RAT} \cdg{STP} \cg{SMCHR} \cdg{UCLID} \cg{Clasp}};
\node[rotate=-60,anchor=west] at (11.5,-0.4) {\scriptsize\tt\cred{Fast SAT Solver}};
\node[rotate=-60,anchor=west] at (11.5,-0.4) {\scriptsize\tt\phantom{Fast SAT Solver} \cg{raSAT}};

\end{tikzpicture}
}
\end{center}
\vskip-10pt
\caption{History of Computer Algebra Systems and SAT/SMT solvers %(not exhaustive; years approximate first release as far as known and as positioning allowed) 
\protect\cite{Abraham2015}
}
\vskip-10pt
\label{fig:tools}
\end{figure}

%%%%% END %%%%%

\section{Satisfiability Checking}
\label{SEC:SAT} 

The \emph{SAT Problem} refers to checking the satisfiability of logical statements over the Booleans.  Initial ideas from Davis and Putnam in 1960 used \emph{resolution} for quantifier elimination; Davis, Logemann and Loveland pursued another line in 1962 with a combination of \emph{enumeration} and \emph{Boolean constraint propagation (BCP)}. A major improvement was achieved in 1999 by Marques-Silva and Sakallah by \emph{combining} the two approaches, leading to \emph{conflict-driven clause-learning} and \emph{non-chronological backtracking}. While the SAT Problem is known to be NP-complete, SAT solvers have been developed which can handle inputs with millions of Boolean variables.  
They are at the heart of industrial techniques for verification and security.

Driven by this success, big efforts were made to enrich propositional SAT-solving for different existentially quantified theories producing \emph{SAT-modulo-theories (SMT) solvers} \cite{BSST09}.  There exist techniques for equality logic with uninterpreted functions, array theory, bit-vector arithmetic and quantifier-free
linear real and integer arithmetic; but the development for quantifier-free non-linear real and integer arithmetic is still in its infancy.  Progress here is required for applications in the automotive and avionic industries \cite{PQR09}.
%\cite{Platzeretal2009}.

SMT solvers typically combine a \emph{SAT solver} with one or more \emph{theory solvers} as illustrated in Figure~\ref{fig:smt}.  
A formula in conjunctive normal form is abstracted to one of pure Boolean propositional logic by replacing each theory constraint by a fresh proposition.
The SAT solver tries to find solutions for this, consulting the theory solver(s) to check the consistency of constraints.
To be \emph{SMT-compliant} the solvers should:
\begin{shortitemize}
\item work \emph{incrementally}, i.e. accept additional constraints and re-check making use of previous results;
\item support \emph{backtracking}, i.e. the removal of previously added constraints;
\item in case of unsatisfiability return an \emph{explanation}, e.g. a small inconsistent subset of constraints.
\end{shortitemize}
%Optimally, they could provide a \emph{satisfying solution} or a \emph{proof of unsatisfiability} also.

\begin{figure}[b]
\begin{center}
\scalebox{0.9}{
\begin{tikzpicture}[font=\large\bfseries, auto, thick, 
box/.style={draw, rounded corners=3pt, inner sep=5pt},
boxA/.style={draw, rectangle split,rectangle split parts=2,text centered,rounded corners=3pt, inner sep=5pt},
boxB/.style={draw, rectangle split, rectangle split horizontal=false,rectangle split parts=2,text centered, rounded corners=3pt, inner sep=5pt}]

\node [box] (ssolver) [color=blue] at (0,0) {SAT solver};
\node [box,color=red] (phi) [above=1.5 of ssolver] {\small input formula in CNF};
\node [box] (eq) [color=blue] at (-3,-1.5) {\small theory constraint set};
\node [box] (ex) [color=blue] at (3,-1.5) {\small \begin{tabular}{c}(partial) SAT or \\ UNSAT $\small \color{blue}+$ \color{blue}explanation\end{tabular}};
\node [box] (tsolver) [color=blue] at (0,-3) {theory solver(s)};

\node [box] (UNSAT) [right=4 of ssolver,color=red] {\small \begin{tabular}{c}SAT or\\UNSAT\end{tabular}};
%\node [box] (SAT) [right=3 of tsolver,color=red] {SAT};

\draw [arrows={-triangle 60},color=red] (ssolver) edge node [above] {\small solution or} node[below] {\small unsatisfiable} (UNSAT);

\draw [arrows={-triangle 60},color=red] (phi) edge node [right, align=left] {\small Boolean abstraction} (ssolver);

\draw [color=blue,-] (ssolver) edge [bend right,out=-20] 
node [above left, xshift=0.8cm,yshift=-0.4cm,fill=white] {\small (partial) solution} 
(eq);
\draw [color=blue,arrows={-triangle 60}] (eq) edge [bend right] (tsolver);

\draw [color=blue,-] (tsolver) edge [bend right] 
%node [right] {\small unsat or partial sat} 
(ex);
\draw [color=blue,arrows={-triangle 60}] (ex) edge [bend right,in=200] (ssolver);

%\draw [arrows={-triangle 60}] (tsolver) edge node [below] {complete sat} (SAT);

\end{tikzpicture}
}
\end{center}
\caption{The typical functioning of SMT solvers}
\vskip-10pt
\label{fig:smt}
\end{figure}

Examples for solvers that are able to cope with linear arithmetic problems are 
\altergo, % \cite{altergo},
\cvc, %~\cite{cvc4}, 
\isat, %~\cite{Article_Fraenzle_HySAT07,ScheiblerKB13},
\mathsat, %~\cite{mathsat}, 
\opensmt, % \cite{Article_Bruttomesso_OpenSMT},
\smtrat, % \cite{smt-rat}, 
\verit, %~\cite{Bouton1}, 
\yices, %~\cite{DM06}, 
and
\zthree. %~\cite{z3}. 
Far fewer tools exist for non-linear arithmetic: 
\isat uses interval constraint propagation;
\minismt tries to reduce problems to linear real arithmetic;
\zthree uses an adaptation of the cylindrical algebraic decomposition (CAD)
method; while
\smtrat
uses solver modules for CAD, virtual substitution, Gr\"obner bases, interval constraint propagation and branch-and-bound. Even fewer SMT solvers are available for non-linear integer arithmetic (undecidable in general).

\section{Challenges and Opportunities}
\label{SEC:Challenges}

SMT solving has its strength in efficient techniques for exploring Boolean structures, learning, combining techniques, and developing dedicated heuristics. Symbolic Computation is strong in providing powerful procedures for sets of arithmetic constraints, and has expertise in simplification and preprocessing.

%For the Satisfiability Checking community to further exploit Symbolic Computation achievements they must be adapted to comply with SMT requirements (CAD, Gr\"{o}bner bases and virtual substitution are tools of particular interest).  
To allow further exploitation by the Satisfiability Checking community, Symbolic Computation tools must first be adapted to comply with SMT requirements (CAD, Gr\"{o}bner bases and virtual substitution are algorithms of particular interest).
However, this is a challenge that requires the expertise of computer algebra developers.  
Conversely, Symbolic Computation could profit from exploiting
successful SMT ideas, like dedicated data structures, sophisticated heuristics, effective learning techniques, and approaches for instrumentality and explanation generation.  Incremental CAD procedures now exist, as do prototypes integrating CDCL-style learning techniques with virtual substitution for linear quantifier elimination.

%\smallskip

We aim to create a new research community \scsc whose members will ultimately be well informed about both fields, and thus able to combine knowledge and techniques to resolve problems (academic and industrial) currently beyond the scope of either individually. To achieve this an EU Horizon 2020 \emph{Coordination and Support Action} (712689) project started in July 2016.  We plan the following actions:

\smallskip

\noindent \textbf{Communication platforms:}  Like Symbolic Computation, Satisfiability Checking is supported by its own conferences (e.g. CADE, IJCAR, SMT) and journals (e.g. JAR); while a role somewhat analogous to SIGSAM is played by the SatLive Forum (\url{http://www.satlive.org/}). 
We have started to initiate joint meetings: 
in 2015 a Dagstuhl Seminar\footnote{\url{http://www.dagstuhl.de/en/program/calendar/semhp/?semnr=15471}} was dedicated to \scsc; 
at ACA 2016 and CASC 2016 there will be \scsc topical sessions; and the first  annual \scsc workshop will take place in affiliation with SYNASC 2016\footnote{\url{http://www.sc-square.org/CSA/workshop1.html}}.  %Other plans include a summer school for young researchers interested in the area.

\smallskip

\noindent \textbf{Research roadmap:} The above platforms will initiate cross-community interactions.  Our long-term objective is to create a research roadmap of opportunities and
challenges; identifying within the problems currently faced in industry, points that can be expected to be solved by the \scsc community.

\smallskip

\noindent \textbf{Standards} 
We aim to create a standard problem specification language for the \scsc community, extending the \smtlib language to handle features needed for Symbolic Computation.  This could serve as a communication protocol for platforms that mix tools; and will be used to develop a set of benchmarks.  
%Standards and benchmarks (and their use in competitions) is thought to have been a great help in the Satisfiability Checking community. 

\end{document}